\documentclass[aps,prl,twocolumn,groupedaddress,nofootinbib,notitlepage,showpacs,floatfix]{revtex4-1}

\usepackage{graphicx,graphics,epsfig,subfigure,times,bm,bbm,amssymb,amsmath,amsfonts,amsthm,mathrsfs,MnSymbol}
\usepackage[matrix,frame,arrow]{xypic}
\usepackage[pdfstartview=FitH]{hyperref}
\usepackage{pifont}
\hypersetup{
    colorlinks=true,       	% false: boxed links; true: colored links
    linkcolor=red,          	% color of internal links
   citecolor=magenta,        % color of links to bibliography
    filecolor=magenta,      	% color of file links
    urlcolor=cyan,           	% color of external links
    runcolor=cyan
}
\usepackage[pdftex]{color}
\usepackage{braket}%Dirac Notation in QM
\usepackage{enumerate}
\usepackage[normalem]{ulem}
\usepackage{subfigure}
\usepackage{overpic}
\usepackage{dsfont}

\usepackage{multirow}
\newcommand{\be}{\begin{equation}}
\newcommand{\ee}{\end{equation}}
\newcommand{\ba}{\begin{eqnarray}}
\newcommand{\ea}{\end{eqnarray}}

\newcommand{\ignore}[1]{}

\newcommand{\Id}{\mathds{1}}

\begin{document}

\title{Variational matrix product operators for the steady state of dissipative quantum systems}

\author{Jian Cui$^{1}$, J. Ignacio Cirac$^{2}$, Mari Carmen  Ba\~nuls $^2$}

\affiliation{$^1$ QOLS, Blackett Laboratory, Imperial College London, SW7 2BW, United Kingdom}

\affiliation{ $^2$ Max-Planck-Institut f\"{u}r Quantenoptik, Hans-Kopfermann-Str. 1, D-85748 Garching, Germany}

\date{\today }

\begin{abstract}
We present a new variational method, based on the matrix product operator (MPO) ansatz, 
for finding the steady state of dissipative quantum chains
governed by master equations of the Lindblad form.
Instead of requiring an accurate representation of the system evolution until the stationary state is attained,
 the algorithm directly targets the final state, thus allowing for
 a faster convergence when the steady state is a MPO with small bond dimension.
Our numerical simulations for several dissipative spin models over a wide range of parameters
illustrate the performance of the method and show that indeed the stationary state is often 
well described by a MPO of very moderate dimensions.
\end{abstract}

\maketitle

{\em Introduction.---}
The physics of quantum systems out of equilibrium poses unsolved fundamental questions, relating to Nature at extreme conditions 
and to the dynamics after long time evolution.
Progress in this field is however hard to achieve, due to the
lack of analytical tools to solve many such problems, and the limitations of existing numerical methods.

In recent times growing attention has been directed to the
out-of-equilibrium physics of open quantum systems, i.e. systems in interaction with an environment.
This interest has been intensified 
by the potential applications to
 the fields of condensed matter physics, statistical physics, and quantum information processing
\cite{Verstraete_QCbydissipation,Zoller_topology,Caizi,Eisert_timing}. 
In particular it has been shown that dissipation can be used to engineer interesting quantum many-body states and to 
perform universal quantum computation \cite{Verstraete_QCbydissipation,diehl08dqc},
ideas which can be explored in the context of 
current experimental setups based on atomic systems \cite{mueller2012review}. 
 A particularly interesting topic is that of dissipative quantum phase transitions (DQPT), namely transitions
 in the non-equilibrium steady state of an open system, 
which may 
arise from the competing effects of the Hamiltonian and the dissipative terms of the 
dynamics. An archetypical example is that of the  model \cite{PhysRev.93.99,carmichael80}, but DQPT have also been
studied in fermionic \cite{prosen2010noise,horstmann2013,PhysRevA.86.013606},
bosonic  \cite{sieberer2013bh}
and quantum spin systems \cite{Prosen_QPTXY2ends,kessler2012,ates2012dyn}.

Finding the stationary state of a generic master equation is not easy, even for  $1D$ systems. 
Analytical treatment is limited to very specific problems, 
such as quadratic fermionic models  \cite{Prosen_3rdquantizationfermi}, 
or systems under special conditions or approximations \cite{degenfeld2014proj,li2014perturbative},
and most often numerical techniques are necessary.

As in the case of pure states an exact numerical treatment is possible only for small systems, due to the exponentially growing computational cost,
which may be even more severe in the case of mixed states.
For pure states, parametrizing the state as a Tensor Network (TN) \cite{cirac2009renorm,verstraete_matrix_2008,Orus2014117}
has proven an efficient alternative, that can successfully capture the physical properties of quantum many body states
in countless situations of interest.
The best example is the tremendous success of the density matrix renormalization group (DMRG) \cite{white92,Schollwock201196},
based on the matrix product state (MPS) ansatz, which provides
a quasi-exact solution for one dimensional problems. 
MPS can accurately describe ground states of gapped local Hamiltonians \cite{verstraete06, hastings07}, 
and methods have been defined to use them also in real time evolution~\cite{vidal_realtimeevolution, daley, hartmann, Banuls}.
In combination with quantum trajectories, the latter have also been applied to dissipative dynamics~\cite{daley09three}.
The natural extension to operators, namely matrix product operators (MPO), can be used as an ansatz for mixed states \cite{Garcia-Ripoll, Zwolak},
which is known to accurately describe thermal equilibrium states for local Hamiltonians \cite{Hastings_thermal,Molnar_thermal}.
Such an extension, in combination with the time evolution algorithms, has allowed the numerical exploration of steady states of spin chains and 
other one dimensional systems under local dissipation (see e.g. \cite{Prosen_MPO2ends,Caizi,Bose_Hubbard_timeevolution,Dephasing_timeevolution,pizorn2013bh,transchel2014tdvp,werner2014puri}).

 This method is formally similar to the search for a ground state using imaginary time evolution \cite{vidal_realtimeevolution}, in that 
a given initial state is evolved until reaching a fixed point of the dynamics. However, 
different to the imaginary time evolution method, 
where the sequence of states visited by the algorithm is not of physical significance,
in the simulation of a master equation, the real evolution needs to be followed, so that
errors in the intermediate state can severely affect the convergence of the procedure.

A better alternative could be given by a variational method, which searches for the null vector of the Lindblad superoperator
within the MPO family, in the spirit of the DMRG variational search.
Such a method would be potentially more efficient than simulating the full evolution, 
specially when the latter
traverses intermediate states with a large bond dimension, but the true steady state is described by a small one,
as is often the case \cite{Caizi,Bose_Hubbard_timeevolution,Bistability_Rydberg}.
In this paper we present such variational method for the steady state of a master equation in Lindblad form.
We illustrate the performance of the algorithm with results for several 
one dimensional models. 
Notice that a variational method, similar in spirit but restricted to density matrices containing only few-body correlations, 
has been recently proposed in \cite{Weimer}.

{\em Basic concepts.---}
\label{sec:basic}
A matrix product state (MPS) for a  quantum system of $N$ $d$-dimensional components, is a state vector of the form 
$|\Psi\rangle=\sum_{\{s_i\}} \mathrm{tr} \left( A_1^{s_1}\ldots  A_N^{s_N}\right )|s_1\ldots s_N\rangle$ \cite{perez2007mps}, 
where each $A_i$ is a $d\times D\times D$ tensor, $D$ is a parameter of the representation called bond dimension, 
and the sum runs over all elements of each individual basis $s_i=1,\ldots d$.
By successively increasing the bond dimension, $D$, the MPS family defines a hierarchy 
of states covering vector space spanned by the tensor product of the individual bases, $\{|s_i\rangle \}$.
The same ansatz can be used to represent operators whose coefficients
 in a tensor product basis have the structure of a matrix product, 
 $\hat{O}=\sum_{\{s_i,r_i\}} \mathrm{tr} \left( A_1^{s_1 r_1}\ldots  A_N^{s_N r_N}\right )|s_1 \ldots s_N\rangle \langle r_1\ldots r_N|$
 These are called matrix 
 product operators (MPO) \cite{Garcia-Ripoll,Zwolak,pirvu2010mpo}.
 The operators can be vectorized using Choi's isomorphism, $|s_i\rangle\langle r_i| \rightarrow |s_i r_i\rangle$,
 which maps any operator $\hat{O}$ to a vector $|\Phi(O)\rangle$, 
 so that it is possible to work in the vector space of operators with the usual MPS techniques.
 
 In order to describe physical mixed states, MPO, or in this case 
 matrix product density operators (MPDO), have to satisfy additional conditions, 
 namely they have to be normalized ($\mathrm{tr}\rho=1$), Hermitian and positive semidefinite.
While the first two conditions are easy to impose on the local tensors of the ansatz, the positivity 
involves the full spectrum of the operator, and is thus a non-local property.
The ansatz can be modified to represent positive operators, 
using a local purification of the state with MPS structure.
In this case, each tensor has a structure
$A_n^{ij}=\sum_k X_n^{ik} \otimes \bar{X}_n^{jk}$,
where the index $k$ sums over the ancillary degree of freedom and the bar indicates complex conjugation.
Although it guarantees positivity, working with the purification ansatz is in general computationally more costly \cite{lubasch2014peps}
and moreover the bond dimension required to write the purification ansatz 
may be much larger than that of the MPO \cite{cuevas2013puri}, so that in practice it is not always the most convenient choice.

{\em A variational search for the steady state.---}
\label{sec:method}
We consider a chain of length $N$, with a quantum system of physical dimension $d$ on each site, and dynamics governed by 
a master equation of Lindblad form, $\frac{d \rho}{d t}=\mathcal{L}[\rho]$, where the rhs is the Lindbladian superoperator,
\begin{equation}
\mathcal{L}[\rho] 
=-i[H,\rho]+\sum_{\alpha}\frac{1}{2} \left(2 L_{\alpha}\rho L_{\alpha}^{\dag}-\{L_{\alpha}^{\dag}L_{\alpha},\rho\} \right).
\label{eq:masterequation}
\end{equation}
The unitary part of the evolution is determined by the system Hamiltonian, $H$. The 
effect of the environment is described by a set of Lindblad operators, $L_{\alpha}$.
 
  The Lindbladian acts linearly on the vectorized $\rho$, as
  \begin{align}
\hat{\cal{L}}=& -i\left ( H\otimes \Id+\Id \otimes H\right) \nonumber \\
&+\sum_{\alpha} \frac{1}{2} \left ( 2 L_{\alpha}\otimes\bar{L}_{\alpha} - L_{\alpha}^{\dag}L_{\alpha}\otimes \Id
 -\Id\otimes L_{\alpha}^{T} \bar{L}_{\alpha}
 \right ).
 \label{eq:Lop}
 \end{align}
 
 The steady state is a fixed point of the evolution, $\frac{d \rho_{\mathrm{s}}}{d t}=0$, and corresponds  to 
 a vector $|\Phi({\rho_{s})}\rangle$ satisfying $\hat{\cal{L}} |\Phi({\rho_s})\rangle=0$, i.e.
 a zero eigenvector of $\hat{\cal{L}}$.
If the Hamiltonian and the individual Lindblad operators have local character, the Lindbladian can be written as a 
 MPO,\footnote{Strictly speaking, it is enough that $H$ and each $L_{\alpha}$ can themselves be written as MPO, which includes short range interactions 
 and dissipation, but can also be applied to approximate power-law decaying terms \cite{pirvu2010mpo}.} and
we can search for the best MPS approximation to its zero eigenvector, which will give us a vectorized MPO approximation for the 
steady state.
Since the operator \eqref{eq:Lop} is not Hermitian, 
in order to use the standard variational search with MPS  
we consider instead the Hermitian product $\hat{\cal{L}}^{\dagger} \hat{\cal{L}}$.
The steady state is also a zero eigenvector of this operator, 
and, since $\hat{\cal{L}}^{\dagger} \hat{\cal{L}} \succeq 0$,
it corresponds to the lowest eigenvalue.
If $\hat{\cal{L}}$ can be written as a MPO, also the product can, 
 and it is then possible to use standard MPS algorithms to approximate its ground state
\cite{verstraete_matrix_2008,Schollwock201196}.
Notice the particular case of Hermitian $L_{\alpha}$ is especially easy,
since the (properly normalized) identity is a steady state, which can be exactly written as a MPS with bond dimension $D=1$.

The fact that we are targeting density matrices requires particular attention,
because not every MPS vector can represent a valid physical state.
The normalization condition $\mathrm{tr}\rho=1$ translates to
$\langle\Phi({\Id})|\Phi({\rho})\rangle=1$, where $|\Phi({\Id})\rangle$ is the (unnormalized) 
vector that corresponds to the trace map, namely the maximally entangled
$|\Phi({\Id})\rangle= \sum_{\{s_i\}} |s_1\ldots s_N\rangle\otimes  |s_1\ldots s_N\rangle$.
A solution which is not orthogonal to this vector can always be normalized to ensure the trace condition.
In general it is more complicated to decide whether a MPS corresponds to a positive operator,
since we do not have access to the full spectrum.
The purification ansatz can guarantee that the search runs over only positive operators, 
but at the expense of more costly local optimizations \cite{lubasch2014peps}.
Hence we use simply the vectorized MPO form and rely on the mathematical 
properties of the problem to provide a physical solution.
Since the evolution generated by $\mathcal{L}$ is a CP map, it must 
have a positive fixed point,
so that if this is non-degenerate, the algorithm should naturally converge towards a MPO
approximation of a positive (and hence Hermitian) operator
\footnote{Strictly speaking, this may fail if the algorithm gets stuck in local minima, or if there are degeneracies, as is known to happen also for DMRG (see e.g. \cite{white98metastable}).}
 and is then expected to be almost positive, with any 
non-positiveness being compatible with the truncation error.

In practice, we find that a suitable warmup phase (see Supplemental Material \cite{supplementary}) allows 
us to avoid solutions with vanishing trace, and improves the convergence of the 
algorithm.\footnote{Notice that it is also possible to use a Lagrange multiplier term, of the form $|\Phi(\Id)\rangle \langle\Phi(\Id)|$,  to favor solutions with  
non-vanishing trace. This can be substracted from $\hat{\cal{L}}^{\dagger} \hat{\cal{L}}$, thus increasing in one unit the bond dimension of the MPO. 
In practice, we found that the warm up phase was enough to obtain physical solutions, without increasing the computational cost. }
Although positivity cannot be checked explicitly (it is in fact a hard problem \cite{kliesch2014mpoNP}), there is a number of 
necessary criteria that any physical state needs to satisfy, such as physically sensible values 
of all single body observables. 
Our algorithm performs a set of such tests and only accepts solutions
that pass them all, otherwise restarting the search with a  different initial guess.
The role of our consistency checks is to discard the least suitable guesses during the warmup phase, 
in order to prepare a suitable initial state for the variational search.
During the later phases of the algorithm, the tests are used as assertions, while we rely on the convergence 
criteria (including that of the effective energy) to stop the calculation. 
After finding an acceptable solution for a given bond dimension, $D$, we 
compute the desired expectation values from the hermitian part of the MPO, $(\rho+\rho^{\dagger})/2$, 
as often done in other algorithms to reduce numerical errors.\footnote{The norm of the non-Hermitian part can also be used as additional consistency check.}
The found solution  is normalized and used as initial guess for a larger bond dimension,
and finally convergence in $D$ is decided when the 
targeted observables are converged to the desired precision.
%\green{Please notice that the role of these consistency checks is merely to discard the most obvious non-physical solutions when D is small, where the positivity of the solution is not guaranteed at all, so as to provide more suitable initial guesses for larger D simulation. In this way, fake increment of bond dimension is greatly depressed. We do not expect these checks to ensure the positivity of the solution. Instead, the positivity is achieved from the convergence of the effective energy towards zero with increasing bond dimension D. Finally we rely on the convergence criteria including that of the effective energy to stop the calculation. }

The algorithm as described here is thus formally equivalent to  the variational ground state search 
for a MPO Hamiltonian over the MPS family, and presents the same scaling, 
 only with $d^2$ playing the role of the physical dimension,
 and with an effective Hamiltonian which has the squared bond dimension of the MPO for $\hat{\cal{L}}$.  
 The gap of  $\hat{\cal{L}}^{\dagger} \hat{\cal{L}}$ will be determinant for the convergence of the algorithm. 
 It is interesting to notice that this is not related to the eigenvalues of
$\hat{\cal{L}}$, but to its (squared) singular values (see \cite{supplementary}).
All in all we find that, for typical cases, the small bond dimension required to approximate the steady state
as a MPO compensates for the additional computational effort associated to $\hat{\cal{L}}^{\dagger} \hat{\cal{L}}$,
provided the Lindbladian is not degenerate. 

Another variational approach has been recently proposed \cite{Weimer}, which chooses to minimize the
trace norm of $\cal{L}[\rho]$. Since this quantity is not efficiently computable, the method in \cite{Weimer} proceeds 
by minimizing an upper-bound to this norm for restricted sets of density matrices.
Our eigenvalue minimization is instead equivalent to finding the vector that minimizes the Euclidean 
 norm $\|\hat{\cal{L}} |\Phi({\rho})\rangle\|$ with the constraint $\| |\Phi({\rho})\rangle\|=1$.
Both minimizations have an exact solution in the physical steady state,
 although they are not equivalent when not exactly on the stationary state.
Using the Euclidean norm of the vectorized expression is preferable in our case,
because the trace norm requires the full diagonalization of the operators, impossible for 
the system sizes we are interested in, while the norm of the vectorized operators is efficiently calculable 
for the MPS ansatz.

{\em Numerical results.---}
\label{sec:results}
To illustrate the performance of the algorithm we apply it to several spin chains, where the unitary and dissipative dynamics show competing effects regarding the coherence of the steady state.

\begin{figure}
\includegraphics[width=.49\columnwidth]{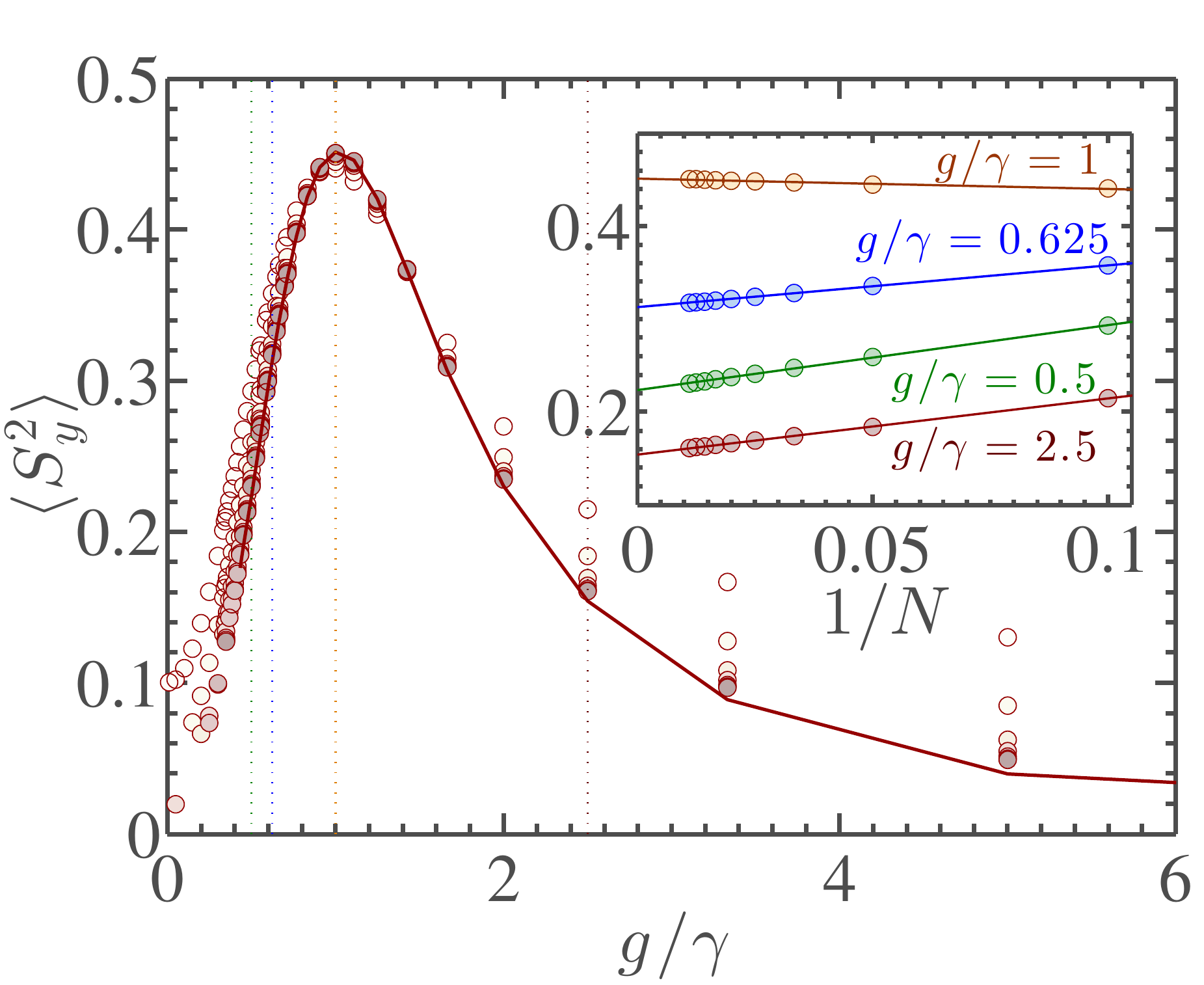}
\includegraphics[width=.49\columnwidth]{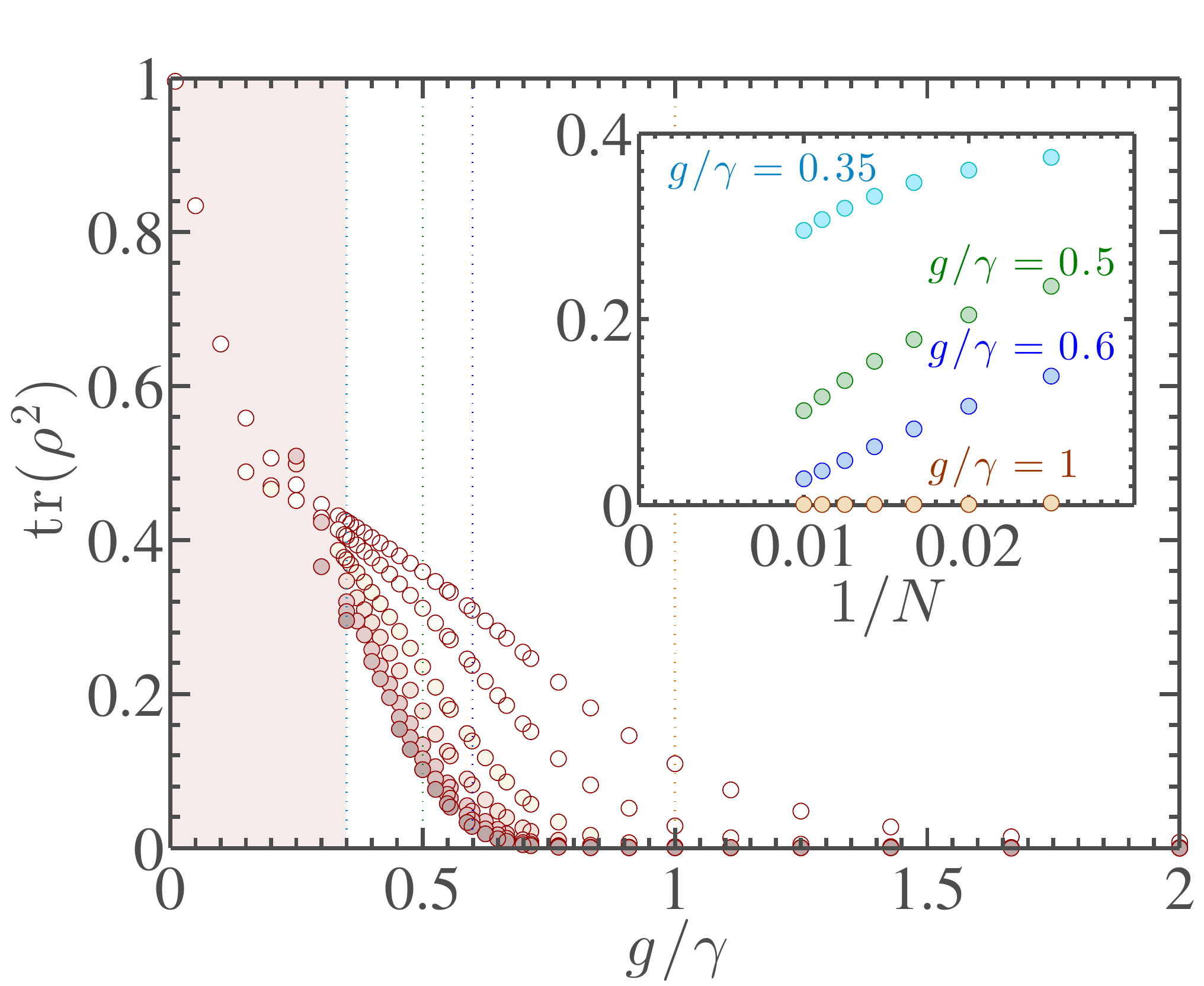}
\caption{Left: Correlation $\langle S_y^2\rangle$ for the low dimensional Dicke model, as a function of $g/\gamma$ for system sizes $N=10-100$ (in increasingly darker shades). The solid line is the result of finite size extrapolation, linear in $1/N$, as explicitly shown in the inset for several values of $g/\gamma$. Right: Purity of the converged steady state  for the same system sizes. For large dissipation (shaded region) we show only results for the smallest systems, for which
the simple variational search converges reliably.
The inset shows the explicit dependence of the purity on the system size for several coupling values.}
\label{fig:dicke}
\end{figure}

\paragraph{Low dimensional Dicke model.}
A typical example of DQPT is exhibited by the Dicke model  \cite{PhysRev.93.99,carmichael80}, in which the collective interaction 
with a single radiation mode induces coherent behavior
on a system of $N$ two-level atoms. 
The regime of parameters required to observe the DQPT is challenging, and the experimental observation of the phase transition has only been achieved recently \cite{baumann2010dicke,hamner2014dicke,Baden2014}.
It is thus interesting to understand the behavior of similar models 
which may then be easier to realize experimentally.
We consider a chain of $N$ two-level systems, where each pair of systems couples coherently to 
a common radiation mode. 
This can be represented by a 
spin-$1/2$ chain, governed by a single-particle Hamiltonian, $H= \sum_{i=1}^N g\sigma_i^x$, and Lindblad operators
$L_{i}=\gamma ( \sigma_{i}^{-}+\sigma_{i+1}^{-})$, for $i=1,\ldots N-1$, instead of the single collective Lindblad operator of the Dicke model,
so that this model can be considered a low-dimensional version of the latter.
We study the nature of the steady state found by the algorithm at varying values of $g/\gamma$ and
increasing system sizes, $N$ up to 100, 
which allows us to perform a finite size extrapolation and study single site observables and correlations in the thermodynamic limit.
In the Dicke model, the superradiant phase transition
(at $g/\gamma^2=N$) \cite{carmichael80} is visible in these observables.
In the low dimensional version we do not find evidence of such transition, although (short range) correlations appear, as shown in figure~\ref{fig:dicke} for $S_y^2=(\sum_i \sigma_i^y/N)^2$.
It is remarkable that for all values $g/\gamma\geq 0.35$ and system sizes $N\leq 100$ the steady state is converged with very small bond dimension, 
$D<30$, most of them even with $D\leq 20$.
At $g=0$ there are however two dark states, namely $|0\rangle^{\otimes N}$
and $(1/\sqrt{N})\sum_{k=1}^N (-1)^k|0^{\otimes(k-1)}10^{\otimes(N-k)}\rangle$. Hence, the
null subspace of $\hat{\mathcal{L}}$
 is four-fold degenerate.
 This hinders the convergence of the algorithm at very small $g/\gamma$, as the steady state is no longer the unique and positive zero eigenvector,
 and the warmup strategy is not enough to guarantee a physical solution, except for the smallest system sizes ($N\leq20$).
  For those converged cases, we can detect 
 the peculiarity of this parameter region
  by analyzing the purity of the solution, shown in figure~\ref{fig:dicke}.
  Indeed, we can find positive solutions with increasing purity, which can be up to $1$ 
  for $g=0$.\footnote{Notice that for $g=0$ any mixture of both dark states will be a steady state, with purity in $[0.5,1]$.}
 In principle one could complement the method with additional techniques to try and select
 the physical steady states (e.g. finding and then processing several orthogonal eigenstates, not necessarily physical) even for larger chains.
 Here we have nevertheless focused on the convergence in the 
 most commonly occurring situation of a unique steady state, 
 where the method can provide the largest gain
 by directly targeting a MPO with small bond dimension.

\begin{figure}
\includegraphics[width=.49\columnwidth]{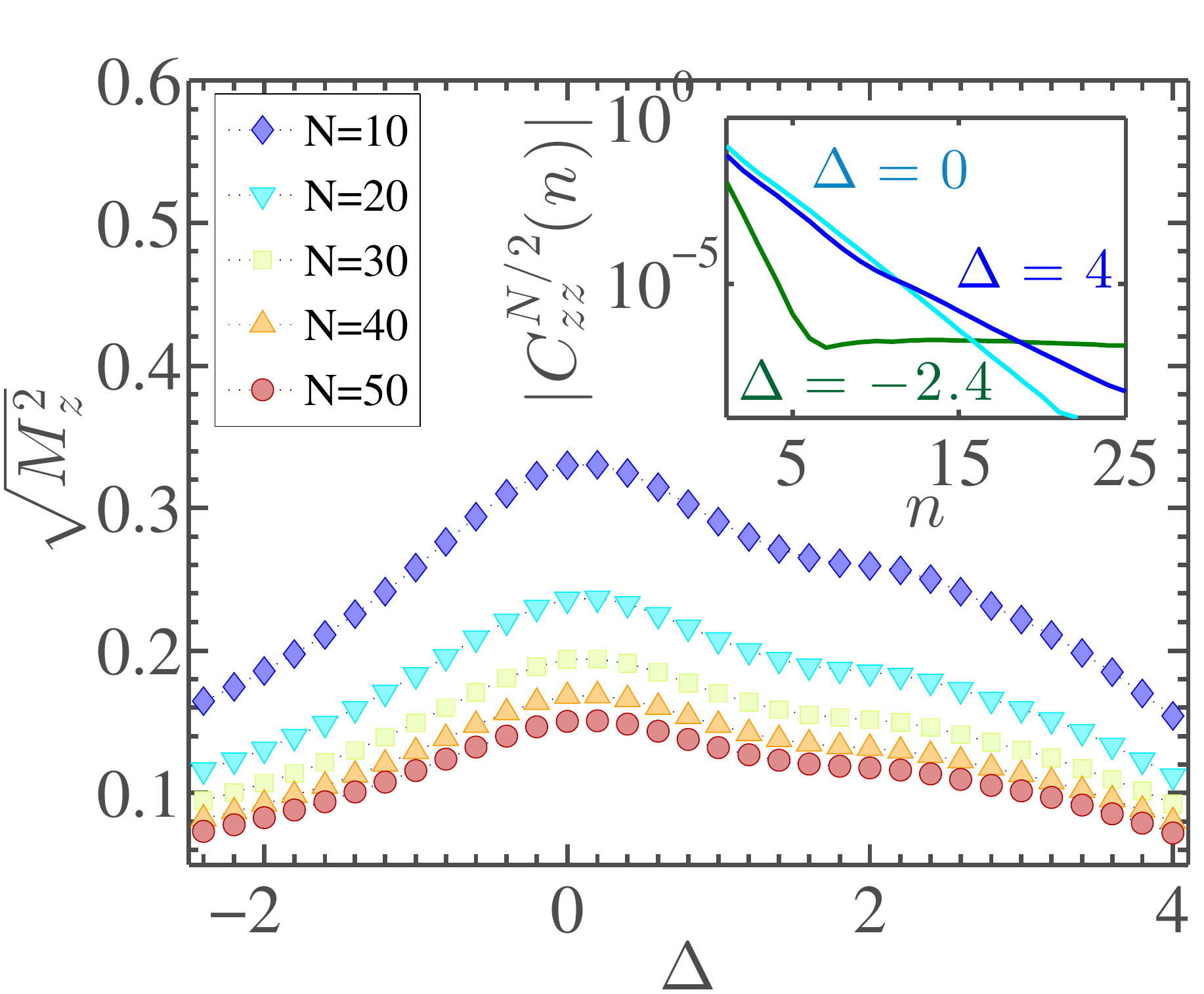}
\includegraphics[width=.49\columnwidth]{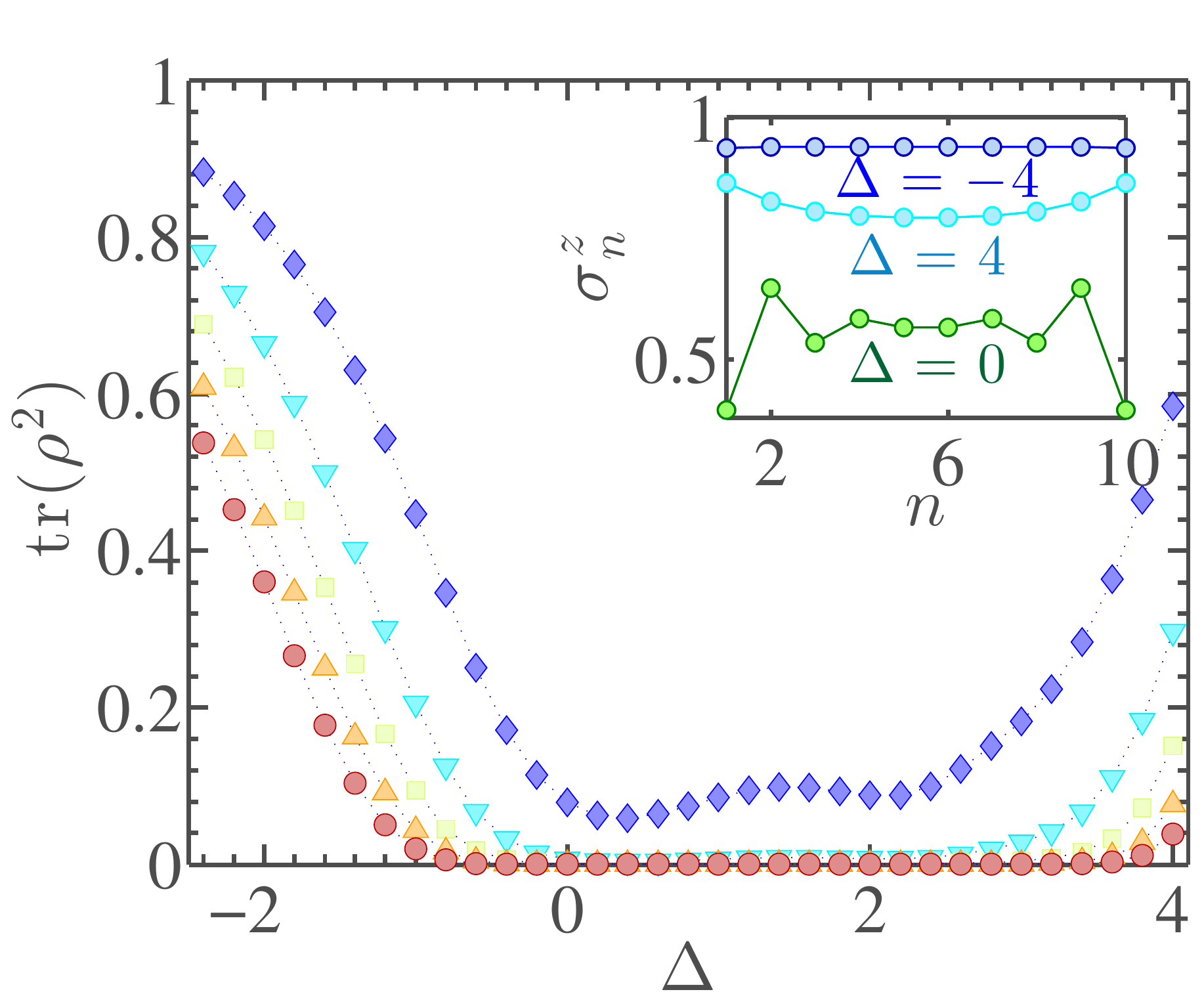}
\caption{Left: antiferromagnetic order parameter $\sqrt{\langle M_z^2\rangle}$ in the steady state of the Ising model with local dissipation, 
for several system sizes, and fixed $\gamma=1$, $V=5$, $\Omega=1.5$, as a function of $\Delta$. 
The inset shows the exponential decay of correlations,
$C_{zz}^{N/2}(n)= \langle\sigma_{N/2}^{z}\sigma_{N/2+n}^{z}\rangle -\langle\sigma_{N/2}^{z}\rangle \langle\sigma_{N/2+n}^{z}\rangle$, 
for $N=50$ in the various regions of $\Delta$.
Right: Purity of the converged steady state for the same system sizes. The inset shows the local polarization, $\sigma_n^z$, for the $N=10$ case. At $\Delta=0$ the antiferromagnetic ordering can be appreciated, while for larger values of $|\Delta|$ the steady state approaches total polarization.
}
\label{fig:ising_loc}
\end{figure}

\paragraph{Dissipative Ising chain.} 
A complementary kind of model is one where the Hamiltonian dynamics induces correlations, for instance an Ising chain, and the dissipation is purely local.
We consider a nearest neighbor Ising interaction, 
$H=\frac{V}{4}\sum_{i<N}\sigma_i^z \sigma_{i+1}^z+\sum_i\left( \frac{\Omega}{2}\sigma_i^x- \frac{V-\Delta}{2}\sigma_i^z\right)+\frac{V}{4}(\sigma_1^z+\sigma_N^z)$, 
and local dissipation given by $L_{i}=\sqrt{\gamma} \sigma_{i}^+$, $i=1,\ldots N$.
Such a model has attracted considerable attention in recent years, as it can be effectively realized in atomic lattice systems 
using Rydberg states \cite{lee2011afm,ates2012dyn,hoening2013crys}. 
In order to compare to existing results in the literature  \cite{lee2011afm}, we fix $\gamma=1$, $V=5$, $\Omega=1.5$.
We compute the steady state for systems up to $N=50$ and study the squared of the staggered magnetization, $M_z=\sum_i(-1)^i \sigma_i^z/N$, equivalent to the 
antiferromagnetic order parameter defined in \cite{lee2011afm}, and the purity of the steady state as a function of $\Delta$ (Fig.~\ref{fig:ising_loc}).
We find that our convergence criteria are met with small bond dimension $D\leq20$. 
Our results show the order parameter vanishing as $N\to\infty$, consistent with short range correlations, which indeed are observed to decay exponentially. 
We observe that the purity of the steady state grows for large absolute values of $\Delta$.
This can be easily understood by going to the interaction picture with respect to the single 
body $\sigma_i^z$ terms in the Hamiltonian.
This does not change the form of the dissipative terms, but
for very large $|V-\Delta|$, the $\sigma_i^x$ terms can be neglected in the rotating wave approximation.
In this situation the single dark state of the dissipation, the fully polarized state $|0\rangle^{\otimes N}$,
is also an eigenstate of the Hamiltonian, and is thus a steady state.

The method can also be applied to other models, for instance with coherence induced by both  the Hamiltonian and the environment \cite{supplementary}.

{\em Conclusion.---}
\label{sec:conclu}
We have presented and analyzed a variational algorithm that searchs for a MPO approximation to 
the steady state of an open quantum system. 
The algorithm is applicable  to any model in which the Hamiltonian and the Lindblad operators can be expressed as MPO.
Instead of simulating the real time evolution of the system, as done by other existing tensor network approaches,
this method directly targets the stationary state, without the need
to precisely describe intermediate states which may need a larger bond dimension than the actual solution.
Thus our technique can allow for a more efficient exploration of the steady state phase diagram.
Our numerical results show that for a variated set of models, with correlations created by the unitary evolution, the dissipation or both \cite{supplementary},
 the steady state is indeed well approximated by a MPO of very small bond dimension,
$D \leq 30$ for sizes up to $N=100$.
This can be directly compared to the bond dimensions required to describe the intermediate states in time evolution methods.
For instance in \cite{hoening2014phd} $D\approx200$ was required for a dissipative Ising chain of length $N=40$.
In \cite{Caizi}, the evolution required $D$ of several hundreds \cite{privatecommunication}, for a XXZ chain of length $N=96$, when the steady state has $D=1$.

Our approach is based on the ground state optimization over MPS for a MPO Hamiltonian,
and relies on the guaranteed existence of a valid, positive steady state.
This basic technique is 
complemented with a warm-up phase or a suitable initial guess, 
 found to be crucial in practice for convergence to a physical result with small bond dimension.

When the steady state is degenerate, the simplest method described in this paper might have problems 
to find a valid guess for the steady state. Specially in the situation of several dark states,  
the null subspace of the Lindbladian contains infinitely many vectors which do not correspond to positive operators and 
hence do not constitute valid physical states. 
\footnote{Notice that this situation could also be adverse for time evolving numerical methods.}
 In principle it would be possible to  complement the current algorithm with additional techniques, 
such as symmetries, in order to reduce the degeneracy, or to construct a candidate steady state from appropriate combinations of 
several linearly independent null vectors, even if non positive.

%%%%%%%%%%

\section{Supplementary}
\subsection{Basics of ``ground state" searching for MPO}
\label{sec:groundstate}
\begin{figure}[h]
\subfigure[The MPO for $\rho$ is mapped to a MPS for $|\Phi(\rho)\rangle$ by Choi's isomorphism.]{
\label{fig:Choi-rho}
   \includegraphics[width=.8\columnwidth]{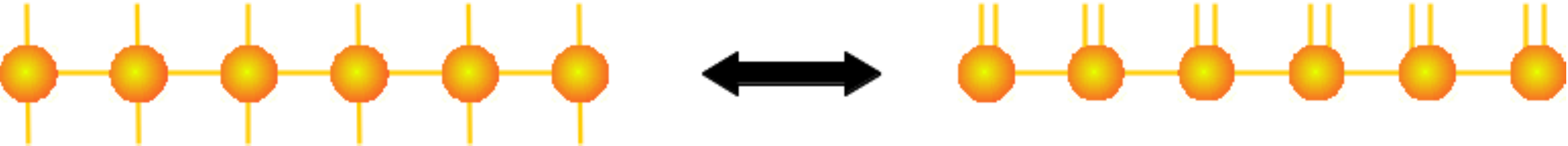}
   }
   \subfigure[For the Hamiltonians and Liouvillian operators considered in this paper, the Lindblad map, $\mathcal{L}(\rho)$ (left),
   is correspondingly mapped to a MPO acting linearly as $\hat{\cal{L}} |\Phi({\rho})\rangle$ (right). ]{
\label{fig:Choi-L}
   \includegraphics[width=.8\columnwidth]{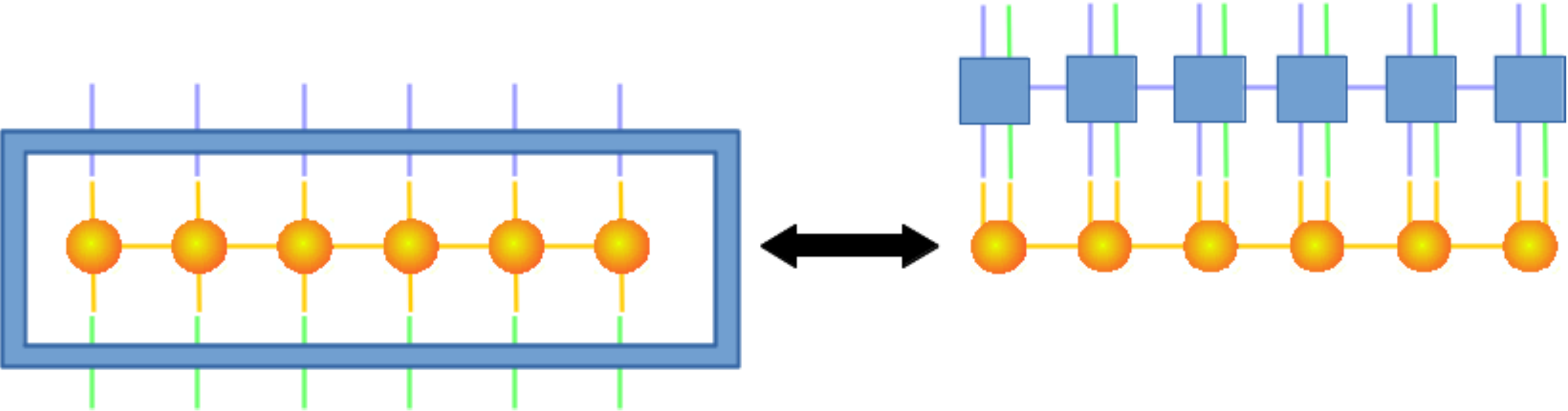}
   }
   \subfigure[The MPO sketched above can be used to compose $\hat{\cal{L}}^{\dagger}\hat{\cal{L}} |\Phi({\rho})\rangle$.]{
\label{fig:LdagLmpo}
   \includegraphics[width=.35\columnwidth]{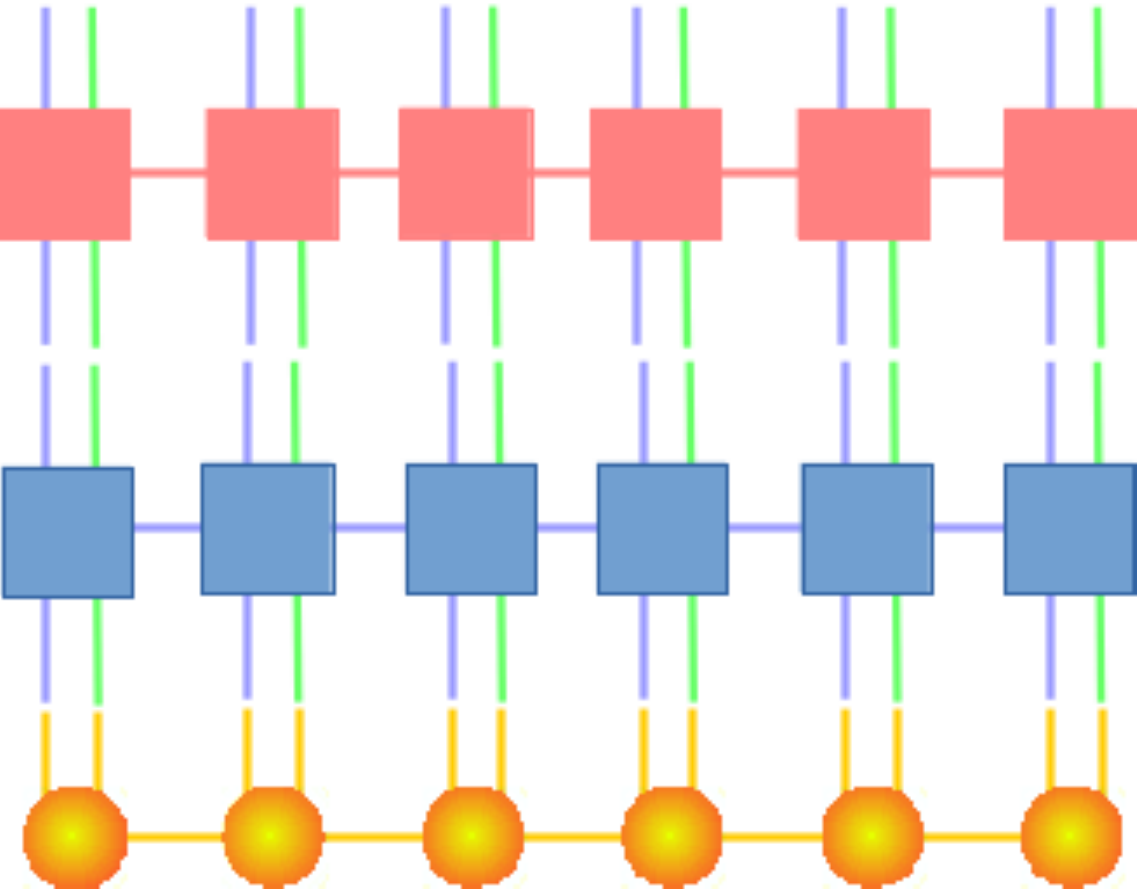}
   }   
      \caption{MPO-MPS representation of the state and the Lindblad super operator.}
   \label{fig:Choi} 
\end{figure}

In the variational search for the best MPS approximation to the ground state of a Hamiltonian, $H$,  
the optimization proceeds using an alternating least squares (ALS) strategy, in which 
the energy, $\langle \Psi|H|\Psi\rangle/\langle\Psi|\Psi\rangle$,
is successively minimized with respect to one of the tensors in the ansatz, while the remaining tensors are fixed
\cite{verstraete04dmrg}.
In the case of the steady state optimization, the basic algorithm is identical, where the product $\hat{\cal{L}}^{\dagger} \hat{\cal{L}}$
plays the role of the Hamiltonian.
We use a MPO description for the superoperator $\hat{\cal{L}}$ (Eq. 2 in the main text),
which can be easily constructed as for the Hamiltonian case,
so that also $\hat{\cal{L}}^{\dagger} \hat{\cal{L}}$ has a MPO representation with the squared bond dimension of the former (see figure~\ref{fig:Choi}). 
In particular, in the cases described in this work, the operator $\hat{\cal{L}}$ contains only local and nearest-neighbor 
terms, but the construction can be extended to more general situations \cite{pirvu2010mpo}.
Running the ALS optimization  requires  the corresponding effective operator
on each tensor, i.e. the contraction of $\langle \Psi|\hat{\cal{L}}^{\dagger} \hat{\cal{L}}|\Psi\rangle$ except for this tensor, 
which can be computed efficiently using the same basic computational routines as the ground state algorithms.

The search for the \emph{ground state} of $\hat{\cal{L}}^{\dagger} \hat{\cal{L}}$
can then produce a MPS approximation to the vector $|\Phi({\rho_{s})}\rangle$,
normalized in Euclidean norm, which minimizes $\|\hat{\cal{L}}|\Phi({\rho_{s})}\rangle\|$.
But not every such approximation will represent a possible physical state,
as those need to be positive operators, with trace one.
Mathematically, there is a zero-eigenstate of $\hat{\cal{L}}$ corresponding to a positive operator, 
so that the algorithm will in general be naturally looking for a physical solution (except in the case, 
illustrated in the main text, of several dark states).
Nevertheless we identify some additional techniques that help ensuring 
the stability and fast convergence of the algorithm.

\subsection{Initial warm-up phase.}
\label{sec:warmup}
In general it is convenient to start the variational search with a small bond dimension, $D$,
and to use the result as initial guess for the search with increased $D$ until the required 
precision is attained.
In the steady state case, nevertheless, if the initial bond dimension is very small, it might happen that the 
algorithm at this stage converges to some vector not corresponding to a physical state,
which constitutes a bad initial guess for the next rounds.
This may lead to very slow convergence, and to an artificial increase of the required 
bond dimension for the final solution.
Hence, for the smallest bond dimension,
instead of directly starting the algorithm on a random vector, we implement a first
warmup phase that constructs a more suitable initial state.
This is then fed to the usual DMRG-like sweep, until convergence.
We also observe that the algorithm behaves better when the bond dimension is increased in small steps.

In particular, for the (close to) reflection symmetric models considered here,
the first sweeps over the MPS tensors are not performed from left to right and back,
as in the regular DMRG algorithms, but symmetrically from the outside in, or from the inside out.
We find that this strategy works best when starting from the separable ansatz, $D=1$, for which 
the algorithm is extremely fast, and can thus be made to converge to numerical precision, and be 
repeated if necessary at a low cost.
If after the algorithm has converged for $D=1$, if the solution is not
compatible with a physical state, the procedure is repeated from a different initial state.
Otherwise, the bond dimension is increased, and the obtained state is used as initial guess.
\footnote{When exploring the parameter space of a certain model, this technique can be used for some parameter values
and then the  converged results can be used as guesses for the model with slightly changed parameter.} 

\subsection{Physical solutions.}
\label{PhysicalSolution}
Checking whether a given MPO corresponds to a physical operator is in general a very hard problem. 
Nevertheless, we can apply some compatibility checks
to discard the most unphysical solutions.
More concretely, for the spin-$1/2$ chains studied, we demand that each of the local magnetizations, $\sigma_n^x$,
$\sigma_n^y$, $\sigma_n^z$, have expectation values within the physical range. 
\footnote{At least within some tolerance, which can be relaxed for the smallest bond dimensions.}
We found this to be enough to identify the most unphysical intermediate solutions, 
and to obtain solutions with non-vanishing trace, so that the 
steady state could be properly normalized,
but it is possible to define additional tests and perform more demanding physicality checks.
Additionally, it might also be interesting to complement the minimization of $\hat{\cal{L}}^{\dagger} \hat{\cal{L}}$
with a Lagrange multiplier term to favor vectors that are not orthogonal to the identity.

\subsection{Deciding convergence.}
\label{convergence}
For the ground state search, convergence of the global algorithm can be decided in terms of the relative variation of the
energy from one value of the bond dimension to the next.
In our case, the exact solution has zero eigenvalue. Instead of its relative variation,
we check the absolute value, so we require that the found eigenvalue is below some threshold, and additionally, 
demand convergence of some physical observables.
In particular, we require that the local polarizations converge, 
by constructing the $N$-component vectors,  $(\sigma_1^{\alpha},\ldots \sigma_N^{\alpha})$, for $\alpha=x,\, y,\, z$ and requiring 
that the relative variation (in Euclidean norm) when increasing the bond dimension  is below $10^{-4}$.
With these criteria, the steady state is converged for all the cases presented in this work.
Additionally we check that other observables (such as the antiferromagnetic order parameter for the Ising model with local dissipation) 
and even the purity, which is far from being a local observable, are converged to a precision better than  $10^{-2}$. 

\subsection{Ising chain with coherent dissipation.}
\label{Ising chain}

\begin{figure}[h]
   \includegraphics[width=.49\columnwidth]{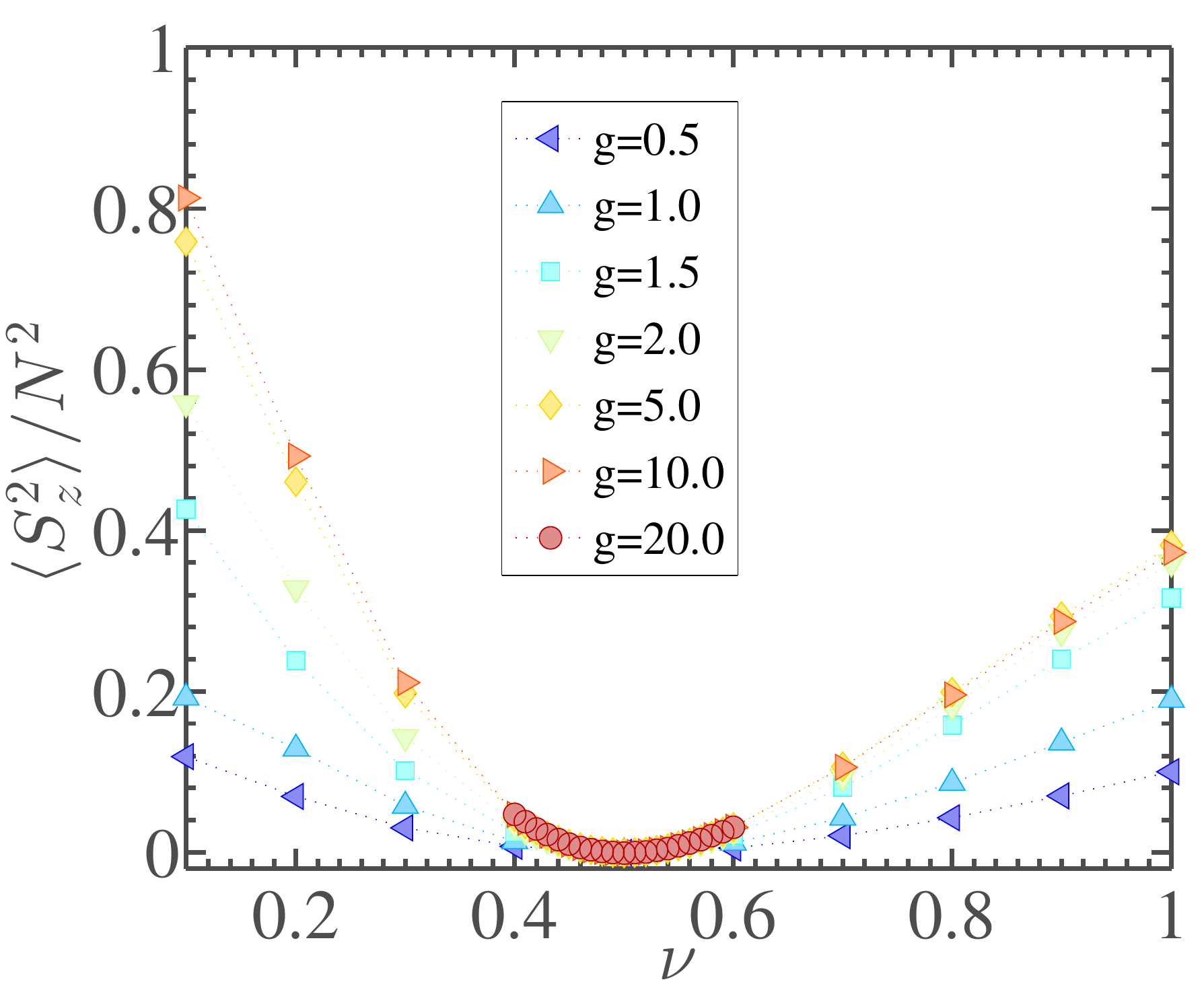}
   \includegraphics[width=.49\columnwidth]{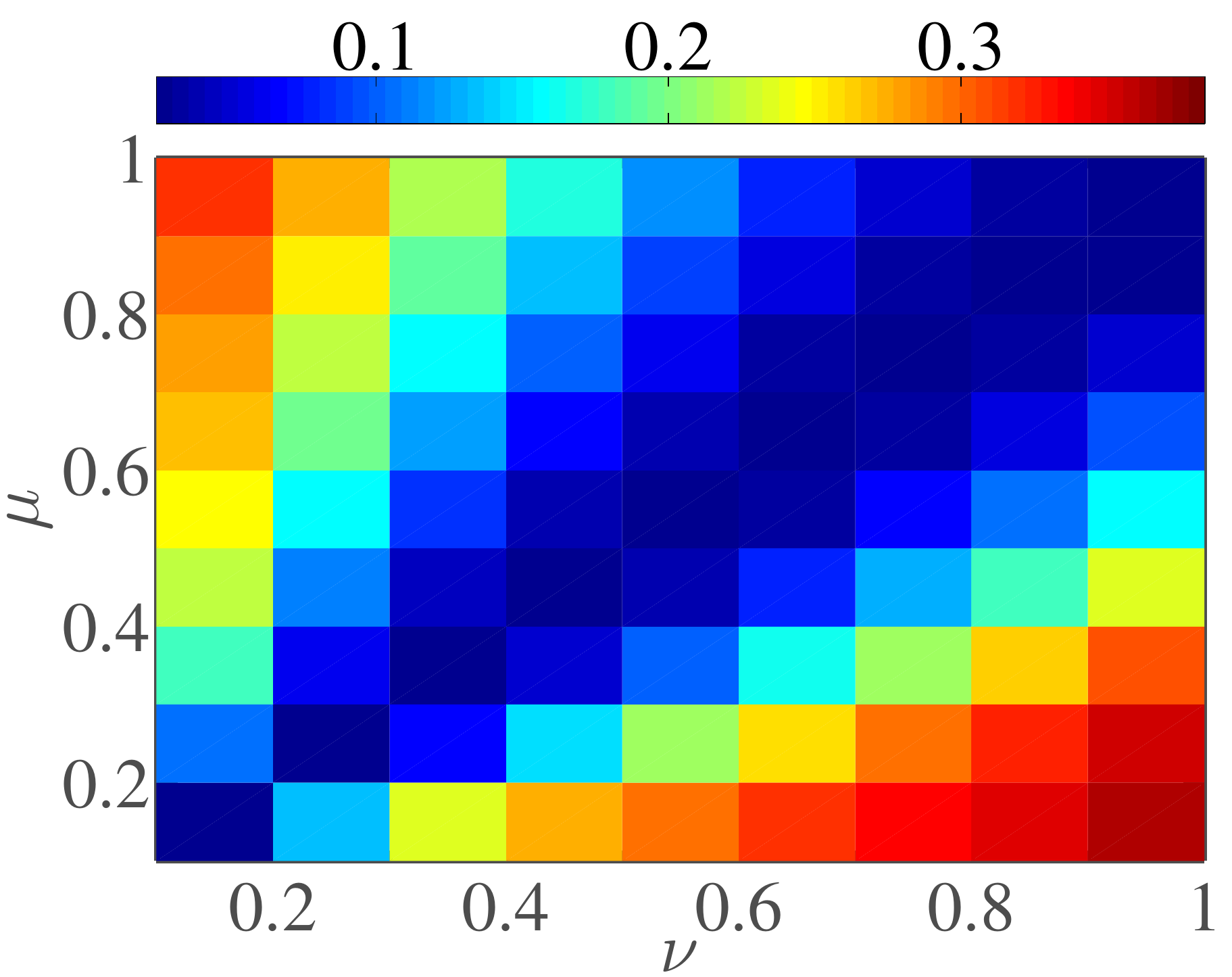}
   \caption{Total $S_z^2$ in the steady state for the Ising model with nearest-neighbor dissipation, as a function of $\nu$ at fixed $\mu=0.5$ for various values of $g$, and after finite size extrapolation from system sizes up to $N=50$.
   The right plot shows the same quantity for varying $\mu$ and fixed $g=1$, for $N=40$.}
   \label{fig:Ising_NN} 
\end{figure}

In the main text we have illustrated the performance of the algorithm in two typical examples, where the Hamiltonian and the dissipation compete to induce or destroy correlations. Here we complement those results with a situation in which both the unitary and the dissipative dynamics can induce coherence. 
Namely, we study an Ising model, 
$H= \sum_i \sigma_i^x \sigma_{i+1}^x + g\sum_i \sigma_i^z$, 
 with nearest neighbor Lindblad operators, 
 $L_i=\mu  \sigma_i^{+}+\nu\sigma_{i+1}^{-}$, $i<N$
 (plus $L_N=\mu  \sigma_N^{+}$).
For chains of up to $N=50$ sites, we find convergence with remarkably small bond dimension, $D<20$. 
Figure~\ref{fig:Ising_NN} illustrates the expectation value of $S_z^2=(\sum_i\sigma_i^z)^2$ in the steady state
for the particular case $\mu=0.5$ and changing values of $g$ and $\nu$, after extrapolating the results to $N\to\infty$.
It is easy to check that in the thermodynamic limit the identity is always a steady state for the symmetric case, $\mu=\nu$.
This explains the vanishing $\langle S_z^2\rangle$ observed in the plot when $\nu=0.5$ for all values of $g$.

\subsection{Spectral gaps of $\hat{\cal{L}}^{\dagger} \hat{\cal{L}}$  and $L$}
\label{sec:gap}
The gap of  $\hat{\cal{L}}^{\dagger} \hat{\cal{L}}$ is determinant for the convergence of the presented algorithm. It plays the same role as the energy gap in conventional DMRG.
For time evolution algorithms the most relevant quantity is the spectral gap of $\cal{L}$, which cannot be related directly to the former. Instead, the eigenvalues of $\hat{\cal{L}}^{\dagger} \hat{\cal{L}}$
are given by the squared singular values of $\cal{L}$.

Here we illustrate how the gaps vary for the two models presented in the main tex, in the case of small chains, using exact diagonalization. 
In principle it would be possible to extend this study and perform a finite size extrapolation, to predict the performance of the algorithm for increasingly larger systems.
Such exhaustive study escapes the scope of this paper, but the finite system results already allow us to appreciate the very different 
situations the algorithm can face.

For the low dimensional Dicke model we observe that the gap closes, even for small systems, in the region of small $g/\gamma$. This is consistent with the 
parameter region on which we found the most difficulties. However, notice that the algorithm easily converged for system sizes $N\approx10$, in spite of the closing gap.

For the Ising chain with local dissipation, when we fix the parameters as in the text and vary only $\Delta$, the gap exhibits much less dramatic dependence. Nevertheless, 
we have shown that in this case, easily treatable by our method, the steady state can present rich features.

\begin{figure}[h]
   \includegraphics[width=.49\columnwidth]{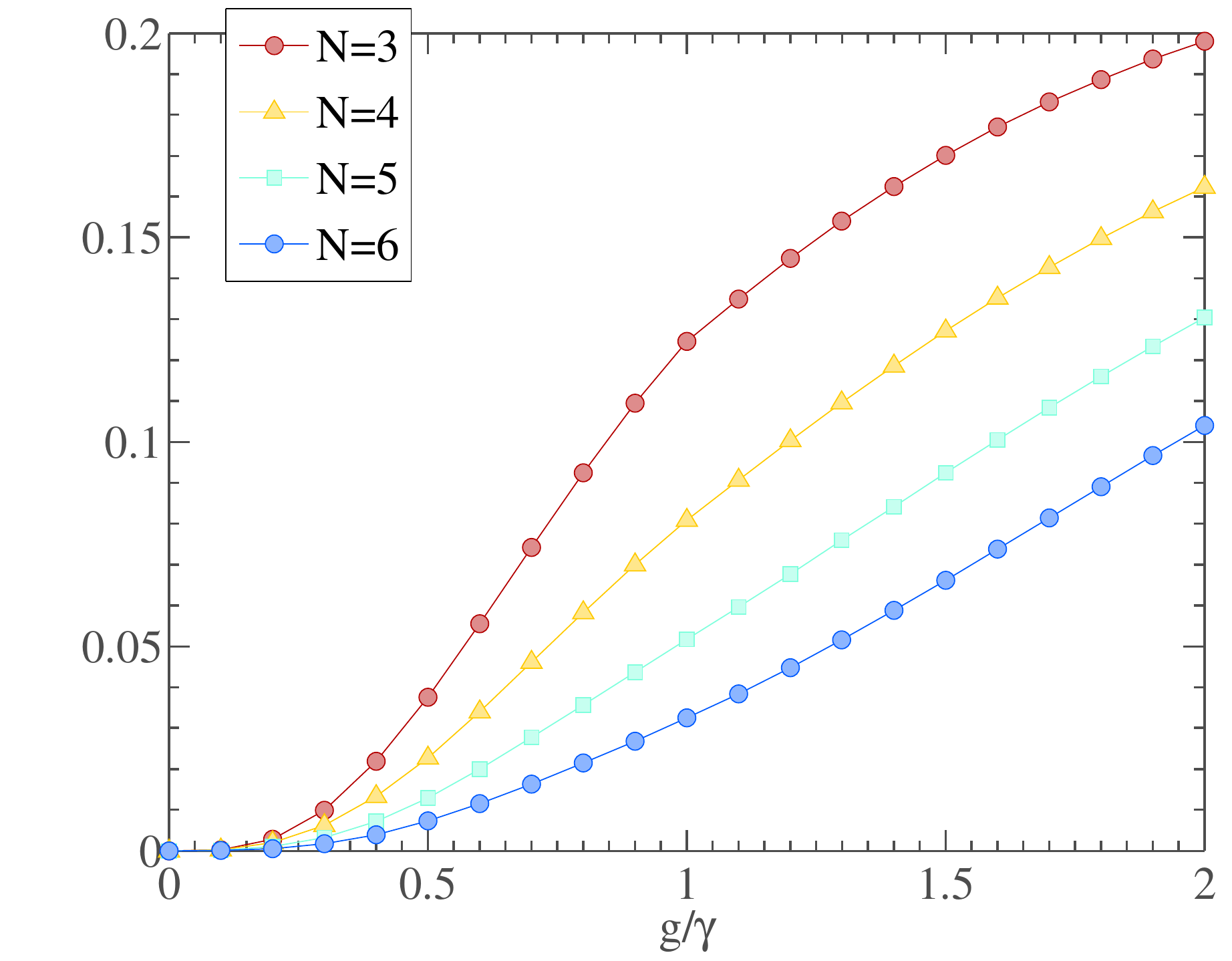}
   \includegraphics[width=.49\columnwidth]{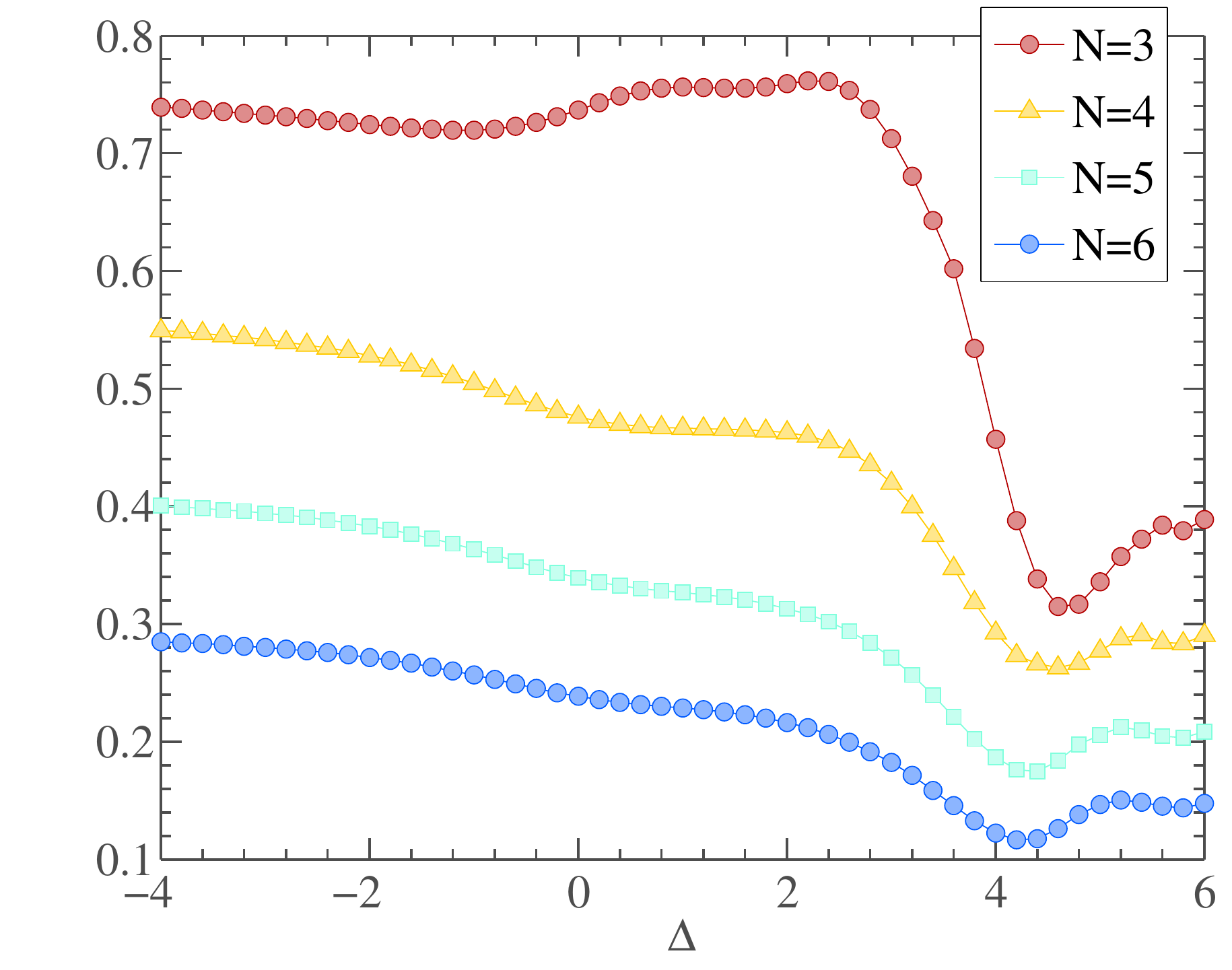}
   \caption{Gap of  $\hat{\cal{L}}^{\dagger} \hat{\cal{L}}$ as a function of the model parameters for small chains, $N=3$--$6$. For the low dimensional Dicke model (left), the gap closes at small values of $g/\gamma$.
   The dissipative Ising chain (right, for $\gamma=1$, $V=5$, $\Omega=1.5$) does not exhibit such a strong dependence of the gap on the free parameter, $\Delta$. In both models the final values $\bra{\Phi(\rho)} \hat{\cal{L}}^{\dagger} \hat{\cal{L}} \ket{\Phi(\rho)} $ that we have accepted are all smaller than $10^{-5}$. }
   \label{fig:gapLLdag} 
\end{figure}

%%%%%%%%%

\begin{acknowledgements}
We are thankful to G. Giedke, D. Porras, H. Weimer, J. von Delft and F. Mintert for discussions. J.C. is supported by the  ``MPG CAS joint doctoral promotion programe(DPP)", and acknowledges Max-Planck-Institute of Quantum Optics, Institute of Physics Chinese Academy of Sciences and Freiburg Institute for Advanced Studies, where part of the research was carried out. This work was partially supported by EU  through SIQS grant (FP7 600645).
\end{acknowledgements}

\bibliography{steadystatereference}

\end{document}